\documentstyle[twoside,fleqn,espcrc2,psfig]{article}

\newcommand{\AmS}{{\protect\the\textfont2
  A\kern-.1667em\lower.5ex\hbox{M}\kern-.125emS}}

\newcommand{\eee}{$e^{\pm}$}

\newcommand{\etal}{{\it et al.}}

\def\astro-d{ {\it ASTRO-D}}

\def\syn{ synchrotron }

\def\fnu{ $\nu \, F_{\nu}$ }

\newcommand{\gamb}{[\gamma_*^2 \, B_{ps}]}
\def\syn{synchrotron }
\newcommand{\be}{\begin{equation}}
\newcommand{\en}{\end{equation}}

\def\bo{  }

\def\rref#1{{\par\noindent \hangindent=3em \hangafter=1
      \advance \rightskip by 0em #1.\par}}
 
\newcommand{\ci}[1]{\cite{#1}}
\newcommand{\bb}[1]{\bibitem{#1}}
 
\newcommand{\loe}{\stackrel{<}{\sim}}
\newcommand{\goe}{\stackrel{>}{\sim}}

\title{
Theory of Gamma-Ray Burst Emission in 
Light of BSAX Results$^{\star}$}

\author{
M. Tavani~$^{a,b}$
\vskip .1in
\noindent
$^a$ IFCTR/CNR, via Bassini 15, I-20133 Milano (Italy).\\
$^b$ Columbia Astrophysics Laboratory, 
Columbia University, New York, NY 10027 (USA).}

\begin{document}

\begin{abstract}
We  briefly discuss the theoretical implications of
recent detections of gamma-ray bursts (GRBs)  by BSAX.
Relativistic shock wave theories of fireball expansion are
challenged by the wealth of X-ray, optical and radio
data obtained after  the discovery of the first 
X-ray GRB afterglow.
BSAX data contribute
to address 
several  issues concerning the  initial and afterglow GRB emission.
The observations  also raise many questions
that are still unsolved.
The synchrotron shock model is in very good agreement with time-resolved
 broad-band spectra (2--500~keV) for the majority
of GRBs detected by BSAX.

\end{abstract}

\maketitle

\section{A Year of Surprise}

The   discovery of X-ray afterglows 
by BSAX revolutionized the
field of GRB research \ci{costa,frontera,piro1}.
The discovery cought us by surprise.
Certainly, there was no theoretical prediction  of
hard \ci{frontera,yoshida2}
 high-energy emission  lasting hours-days after
the main events. `Pre-BSAX models' of GRB shock waves
predicted a much faster decay of the hard component
produced by the initial impulsive particle acceleration 
(e.g., \ci{mesz4}).
Before the BSAX discovery, the only hope to
detect GRB counterparts  was believed to be 
searching for  rapid UV/optical transients lasting
a few minutes/hours or possibly delayed radio flares
(e.g.,~\ci{rhoads}).

The ingenuity of the BSAX team was remarkable in
 ignoring theoretical models and carrying out the fastest
ever slews of an X-ray satellite to GRB error boxes.
BSAX discovery shattered old beliefs, and opened a 
new way of confronting difficult problems of GRB physics.
A more  complete picture of the GRB
phenomenon is now emerging with all its complexities and
puzzles.
All theoretical  models are challenged by the wealth of
X-ray, 
------------------------------------------------------\\
\noindent
{\small ($\star$) Paper presented at the Symposium 
{\it The Active X-Ray Sky: Results from Beppo-SAX and Rossi-XTE},
Rome (Italy), 21-24 October 1997, Accademia Nazionale dei Lincei.
To be published in Nuclear Physics B Proceedings
Supplement, eds. L. Scarsi, H. Bradt, P. Giommi \& F. Fiore.
}

\noindent
optical and radio
data, and many problems remain unsolved
at the moment. We briefly discuss\\
 here  some of the open 
issues.

BSAX showed for the first time that a substantial fraction of
GRB energy is dissipated in the X-ray range
 at late times (hours, days, sometimes
weeks) after the main impulsive events.
{\it Are classical GRBs the tip of the iceberg of  more
complex radiation processes  lasting 
much  longer  than  expected from the  cooling timescales
of initial pulses  ?}
The `delayed' gamma-ray emission discovered
by EGRET in the case of GRB~940217 \ci{hurley}  clearly shows how
 GRB `durations'  determined by BATSE in the 50--300~keV
range can be misleading. 
The delayed gamma-ray emission of GRB~940217 can now 
naturally be interpreted as a manifestation of GRB afterglows.
Particles  associated with this type of  bursts
can remain accelerated for a timescale $\sim 10^2-10^3$ times
longer than the decay timescales of initial GRB pulses.
The fluence of the delayed gamma-ray emission of GRB~940217
is larger than $ 10\%$ of that detected for the main event \ci{hurley}.
Also in the case of  X-ray afterglows detected by BSAX,
at least a fraction  larger than $ 10\%$ of the total fluence
is emitted at late times (e.g., \ci{costa,piro2}).

The total energy inferred from the X-ray afterglow 
\ci{piro2} and possible optical 
counterpart \ci{pian} of GRB~970508 ($\goe  10^{52}$~ergs)
 exceeds the most
optimistic estimates
of coalescing neutron star models (e.g., \ci{paczinski0}).
{\it If GRB sources are at extragalactic distances, do they imply  a
new kind of explosive phenomenon ?} (see also the speculations
of ref.  \ci{paczinskinew}
to be compared with those of  ref. \ci{paczinski0}).

Renormalizing the logN--logP distribution of GRBs 
(see Fig.~1) in terms of
redshift for an assumed luminosity function  can be done
for GRB~970508  associated with 
the (lower limit) optical transient redshift $z=0.83$ \ci{metzger1}.
Unless GRB~970508  is anomalous,
it can be shown that the standard candle assumption
cannot satisfactorily describe the GRB brightness distribution, contrary 
to the conclusions of previous studies (e.g., \ci{bloom}).
Cosmological models of GRBs have to be formulated now
without the appeal of a natural energy scale provided
by the neutron star coalescence model.
{\it What is the origin of the spread in luminosity
(and spectral  characteristics as shown in ref. \ci{tavani-logn})
of GRB sources ?}

BSAX discoveries opened the way for rapid follow-up
observations in the optical and radio bands.
At present, four  out of eight  GRBs detected by BSAX
(GRB~970228, 970402, 970508, 971214) unambiguously show
the existence of 
fading X-ray sources within the WFC error boxes ($10-30 \rm \, arcmin^2$)
pointed $\sim 8 $~hours after the events. 
Of the remaining four  GRBs, 
two error boxes  could not be rapidly pointed
(GRB~960720 and 980109),
 and the others  pointed  within
$\sim 14-16 $~hours show faint sources with ambiguous associations
(GRB~970111 and 971227).
Another GRB error box
pointed by ASCA within 1 day after the event (GRB~970828)
also shows
 an X-ray fading afterglow source \ci{murakami,yoshida2}.
Usually the X-ray flux decay is well represented by a power-law
of the type $t^{-\alpha}$.
{\it Why are the time exponents $\alpha$'s  of X-ray afterglows  different
from burst to burst~? 
Is there any correlation between the  afterglow strength
and the initial GRB peak intensity or spectrum  ?}

At present, three  optical transients were identified within the
WFC error boxes of GRB~970228 \ci{vanparadijs},
 GRB~970508 \ci{bond} and GRB~971214 \ci{halpern}.
All 
are within the error boxes of the fading X-ray sources,
reinforcing their associations with their respective GRBs.
However, they all show different characteristics.
%
The delayed ($\sim 2 $~days) 
optical transient (OT) associated with
GRB~970508 resulted in the spectral identification of
absorption lines of an object at $z \simeq 0.83$
(either the host  or a foreground galaxy \ci{metzger1}).
After the delayed rise, the optical lightcurve follows a
power-law decay  of index $\sim 1.17\pm 0.04$
for several weeks (e.g., \ci{pian}).
The nature of the OT associated with GRB~970228 is
currently controversial, with a  nebulosity
 near a fading pointlike  OT \ci{vanparadijs,sahu,caraveo,fruchter}.
The  OT  was still  detectable
by HST $\sim 6$~months after the event near $V\simeq 28$
\ci{fruchter}. Its inferred optical lightcurve shows
clear deviations from a power-law decay 
of index $\sim 1.14\pm 0.05$ \ci{fruchter}. 
{\bo
The OT associated with GRB~971214  shows an
initial   power-law decay of index $\sim 1.4\pm 0.2$ \ci{halpern},
 and a possible  flattening near $R = 25.6$ 
about $10$~days
after the event \ci{kulkarni}. 
{\it What is the nature of the faint nebulosities
associated with GRB~970228 and GRB~971214 ?
If these are distant star-forming galaxies,
why would GRBs be preferentially hosted near the cores
of these galaxies rather than farther away as coalescing
neutron star models predict ?
If the comoving GRB formation rate follows that of
star forming galaxies (strongly peaked near $z\simeq 1$
\ci{madau}), why are only the bright, harder and longer
GRBs showing the strongest deviation from the Euclidean
brightness distribution  \ci{tavani-logn} ?}
Significant luminosity and spectral cosmological evolution
of GRBs at $z\goe 1$ may be required \ci{tavani-logn}.

\begin{figure*}  
 \vspace*{-3.cm}
\centerline{\psfig{file=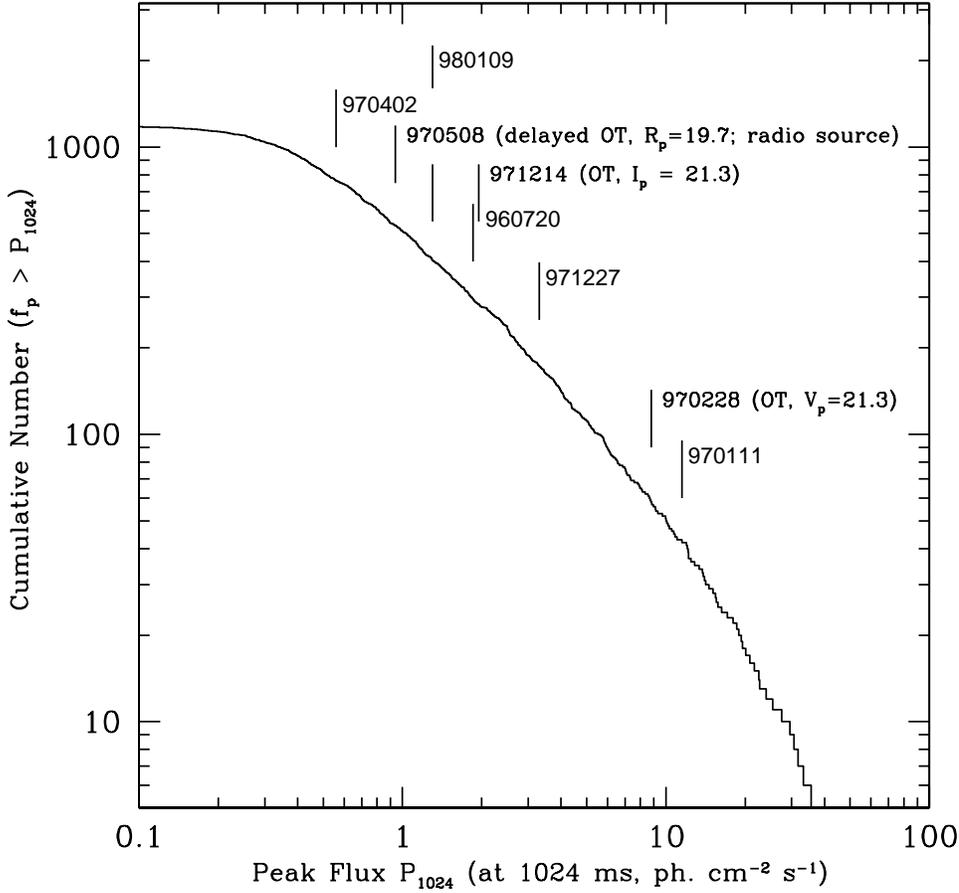,height=13.cm,width=13.cm,clip=}}
 \vspace*{-.5cm}
\caption[image]{
GRBs detected by BSAX marked on the cumulative
brightness distribution
of the 4th BATSE catalogue \ci{paciesas}.
Peak intensities in the BSAX-GRBM detector have been rescaled
to fit BATSE's energy range (50-300~keV) assuming a power-law
photon index of 2.
BSAX WFCs are clearly  capable of detecting  faint GRBs near
the detection threshold for BATSE.
Three GRBs associated with optical transients (OT) are
marked with their respective peak magnitudes 
(GRB~970228 \ci{vanparadijs},
GRB~970508 \ci{castro},
GRB~971214 \ci{halpern}).
No OT was discovered in other GRB error boxes.
The only GRB associated with a
(possibly scintillating)  radio source is 
 GRB~970508 \ci{frail} of low gamma-ray flux \ci{piro2}. 
Radio searches  in other GRB error boxes were rapidly
performed with null results \ci{frail3}.}
\label{fig1}
\end{figure*}

About half of the GRBs promptly studied by optical searches
do not show any OT below  $R\sim 22-23$ (GRB~970111,
970402, 970828, 970828, 971227, 980109).
{\it Is the lack of OTs in more than half of GRBs 
due to absorption within the host~?
Is the lack of detectable absorption in many of the X-ray
afterglow spectra obtained by SAX consistent with
optical absorption [taking into account redshifting  by
$(1+z)^3$ of the X-ray cutoff energy]~?}

GRB~970508, the event associated with an extragalactic OT,
is `anomalous' in many ways when compared with other GRBs.
{\bf  (i)} It is the only GRB with a non-monotonic X-ray afterglow
decay $\sim 1-3$ days after the event \ci{piro2}.
Four BSAX TOO observations were necessary to finally observe  its
decay in the 2--10~keV band (but apparently not in the softer
LECS band \ci{piro2}).
{\bf  (ii)} Its associated OT was detected to rise $\sim 2$~days after
the event after a stable plateau state \ci{bond},
contrary to the other two GRBs showing monotonically decreasing
 OTs (GRB~970228 \ci{vanparadijs}, and GRB~971214 \ci{halpern}).
{\bf  (iii)} GRB~970508 is also the  only GRB so far associated with a
(scintillating~?) radio source \ci{frail}.
{\it Are the peculiar properties of GRB~970508 
caused by the `environment' surrounding the GRB source, or can
they be attributed to a background AGN ?
Why is  only the very weak GRB~970508 associated with an
apparently persistent \ci{frail2}
radio source~?}



 If delayed radio emission is common in
bright GRBs produced by relativistic fireballs
(e.g., \ci{rhoads}), more radio detections would have been
expected (see~Fig.~\ref{fig1}).

Finding  satisfactory explanations to all these questions
will be challenging for any theory of GRBs.
The  issues facing cosmological models 
\ci{katzpiran,tavani,vietri,waxman,wijers}
in the `post--BSAX era'
of GRB research include:
an energy crisis, burst number density evolution vs. star forming
galaxies,  luminosity and spectral evolution,
absorption and reprocessing properties of GRB environments in 
distant galaxies, diversity of X-ray and optical decays, 
persistent optical emission $\sim 6$~months after GRB~970228,
and lack of radio emission for the majority of GRBs.

Other models should be considered, and a superposition of
GRB populations of different origins may still be viable.
More detections of optical/radio  transients in
GRB error boxes  are definitely needed.

\section{Contributions by BSAX}

BSAX observations  contribute to 
address  important aspects of the GRB emission mechanism
for both the prompt impulsive emission and the
delayed afterglows. We indicate here ten important
contributions by BSAX that can be used for detailed
 theoretical modelling.

{\bf (1) } {\it Extending the GRB spectrum to X-ray energies:
time resolved spectroscopy.}
BSAX data complement and improve those obtained by GINGA
\ci{yoshida,strohmayer}. Combining  WFC and GRBM data for
sufficiently intense GRBs can provide a unique database for
studying time-resolved spectroscopy from $\sim 2$~keV to
hundreds of ~keV.
No systematic low-energy `suppression' or spectral
`up-turns' is detected by BSAX, clearly confirming that
possible deviations  \ci{preece2}
from the phenomenological Band's model
\ci{band} affect a minority of GRBs.

{\bf (2)}
{\it Discovery of substantial spectral re-hardening
(e.g., GRB 970228) for late GRB pulses.}
Usually hard-to-soft  spectral evolution  
dominate across individual
GRB pulses \ci{norris2} or among different pulses of complex GRBs
\ci{pendleton}.
The broad-band detection by BSAX of the complex GRB~970228
\ci{frontera} clearly shows how a second train of pulses
separated from the first part of the GRB by several tens of
seconds can be much harder than expected from simple extrapolation. 
Particle  re-energization clearly occurs tens of seconds after
the main event, giving support to the idea that the second train
of pulses in GRB~970228 may be related to its strong X-ray afterglow
\ci{costa}.

{\bf (3)}
{\it Confirming the energy dependence of GRB lightcurves.}
Many initial GRB pulses  detected by BSAX show a characteristic
apparent `time delay'  between the  hard (50-600~keV)  and soft
(2-30~keV) lightcurves. This phenomenon has also been clearly observed
in different energy ranges by CGRO instruments (e.g., \ci{norris2}).
Rapid spectral evolution with the `sweeping' of a fixed energy
band by the peak energy  $E_p$ of the \fnu spectrum can 
satisfactorily explain the observations.

{\bf (4)} 
{\it Pulse width vs. photon energy relation:
$  \Delta \, \tau  \propto E^{-1/2}$}.
BSAX clearly confirms the pulse broadening as a function of
decreasing photon energy detected also in the BATSE  range 
(e.g., \ci{norris2}).
This feature of GRB pulses strongly support
synchrotron models of emission \ci{ta1}.

{\bf (5)}
{\it Discovery of long-duration X-ray afterglows  (XRA)
occasionally lasting 
$\sim $~days/weeks after the events (as in the case of
GRB~970228 \ci{costa}).}
Lower limits to the X-ray afterglow fluences  can be
deduced in the range of several tens of a percent
of the main pulse fluences. 

{\bf (6)}
{\it Discovery  of `average' XRA power-law decays:
 $  F_x \sim t^{- \alpha}$, with $  \alpha = 1.1-1.6$.}
Interpreted as manifestations of persistently accelerated
particles, XRAs can be used to derive constraints on the
initial and maximum energies of radiating leptons in 
relativistic  shock waves \ci{tavani}.
Only forward shocks agree with BSAX observations
\ci{tavani,wijers}, predicting an X-ray flux decay of the type
$  \bf f_x  \propto  \xi_x (t) \, t^{-3/2} $ 
with $\xi_x (t)$ an X-ray  `window' function depending on
spectral evolution. The requirements on the initial
($\gamma_*$) and maximum ($\gamma_m$)  energies of particles
accelerated by an impulsive  relativistic shock  are
stringent  \ci{tavani}
\be \gamma_{m}^2 \, \gamma_*^{1/2} \simeq 10^{15} 
\, n_o^{-1/2} \, \Gamma_{0,2}^{-3/2} \, \lambda_B^{-1/2} \,
(t/{\rm 8 \; hr})^{3/2}  \en
where
$n_o$ is the average number density of the surrounding medium,
$\Gamma_{0,2}$ the initial value of the bulk Lorentz
factor of the shock wave in units of $10^2$,
$\lambda_B$ a parameter set to unity for equipartition
between kinetic and  magnetic field energy densities,
and $t$ the time after the burst in the observer frame.
Fluctuations are occasionally detected in BSAX X-ray afterglows
possibly indicating re-acceleration episodes
 (see also the  GRB~970828 afterglow detected by ASCA \ci{yoshida2}).

{\bf (7)}
{\it Discovery of strong violations of power-law XRA decays:
the case of GRB~970508 \ci{piro2}}.

{\bf (8)}
{\it First observation of a non-thermal XRA spectrum (GRB~970228)
of photon index near 2  \ci{frontera}.}
This important discovery
is also supported by a similar 
  spectral determination of the 
GRB~970828 afterglow by ASCA \ci{yoshida2}.
It can be shown that the combination of flux decay and
spectral properties of GRB afterglows is  inconsistent with
simple cooling models of neutron star surfaces \ci{tavani}.

{\bf (9)}
{\it Lack of prominent X-ray `precursors'.}
WFC data can be used to test for the first time the possible
existence of precursor X-ray emission as predicted in
several fireball models.
No evidence for fireball thinning is found,
implying values of the bulk Lorentz factor of the 
expanding shell $\Gamma \goe$~a few.

{\bf (10)}
{\it Delayed GRB emission allows optical identification}.
Probably the most important consequence of the observations
originating from the BSAX detections.

\section{The Synchrotron Shock Emission Model}

Among the many possible  spectral models of GRB emission
(Compton attenuation due to  Klein-Nishina scattering effects
\ci{brainerd}, absorption effects by thick material near compact 
objects \ci{liang3}, or processes leading to
 electron/positron creation and beaming \cite{baring}),
the shock synchrotron model (SSM) is of special relevance
to relativistic fireball models \ci{ta1,ta2}.
BSAX results  confirm  SSM predictions to $\sim$~keV energies
for the majority of detected GRB pulses (exceptions  will be
discussed below).
The SSM
 is based on optically-thin \syn emission of relativistic particles
(electrons and/or \eee-pairs) radiating in the presence of a weak to moderate
magnetic field (to avoid magnetic absorption processes) \ci{ta1,ta2}. A model
that successfully describes broad-band  GRB spectra is based on a
particle energy distribution 
consisting of  a relativistic Maxwellian
below a critical energy. This
reflects the physical conditions of a thermalized
 electromagnetic/particle outflow  produced by
an initial impulsive burst.
 Depending on the
characteristics of external media and their radiative environments, an MHD
wind is assumed to interact in an optically-thin environment with magnetic
turbulence or  hydromagnetic shocks leading  to
rapid  particle acceleration and
to the formation of a prominent supra-thermal component of power-law index
$\delta$.
Depending on the efficiency of the acceleration mechanism,
the particle energy distribution can have different shapes \ci{ta2}.
The SSM results in the  dimensionless spectral function 
of refs. \ci{ta1,ta2}.
Fig.~2 shows SSM calculations of GRB spectra that reproduce
with good accuracy the broad-band spectra obtained by CGRO.
BSAX spectral data allow to extend the energy range of these
fits to $\sim$~keV energies  with good agreement between
theory and observations.
We note that the phenomenological spectral model by Band \ci{band}
is in excellent agreement with SSM spectra, that therefore
provide a natural theoretical foundation for it.

\begin{figure*}
\vspace*{-4.cm}
\centerline{
\psfig{file=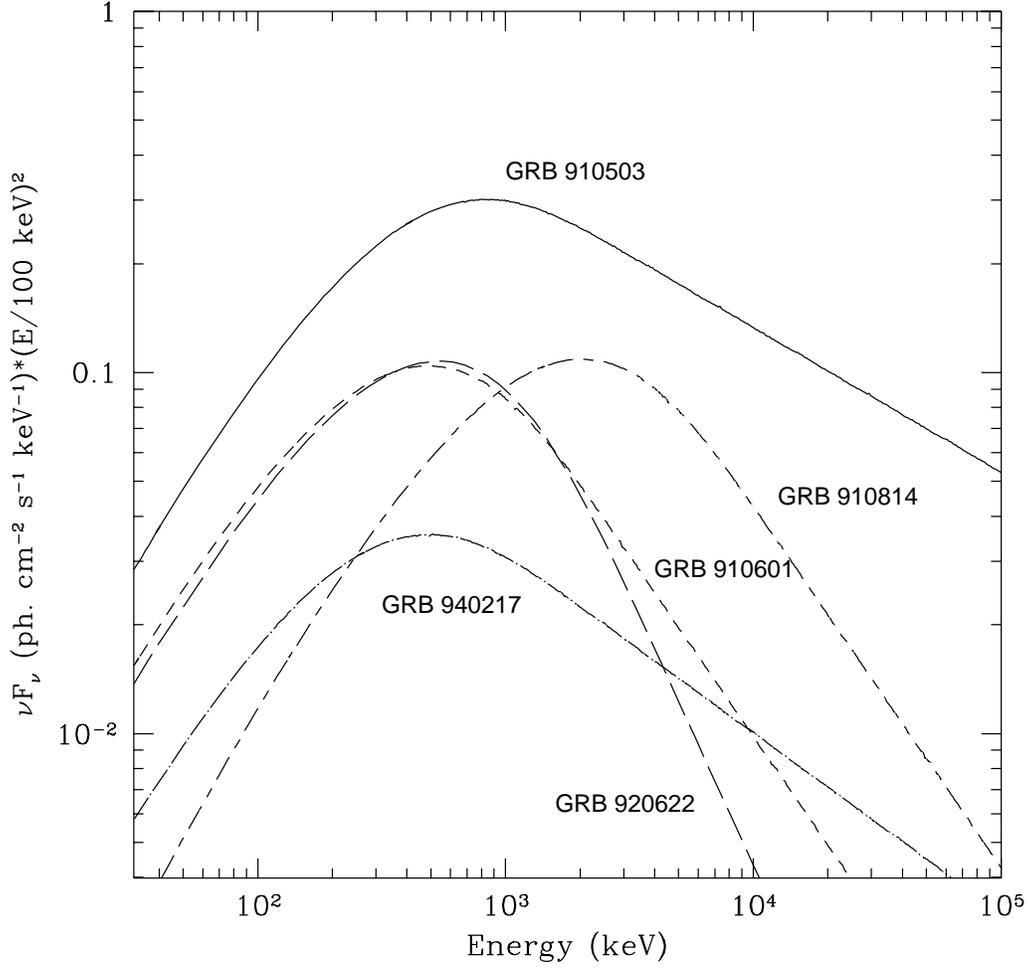,height=14cm,width=14cm,clip=}
}
\caption{
Calculated SSM \fnu spectra reproducing the burst-averaged
spectra of  the GRBs simultaneously detected by BATSE, COMPTEL
and EGRET on board of CGRO (adapted from ref. \ci{ta1}).
Also time-resolved BSAX spectral data are in agreement 
with these SSM calculations assuming optically thin radiation
environments.
BSAX data are consistent with SSM spectra
 of  initial peak energies in the range 
$10~{\rm keV}
 \loe  E_p \loe 500$~keV and subsequent hard-to-soft spectral evolution.
An interesting exception is indicated by
 the initial part of the GRB~960720 pulse showing an apparent absorption
below $E_p \simeq 100$~keV compared to the SSM prediction.
Most likely, a modification of optically thin SSM by relativistic
plasma effects is implied for this type of bursts \ci{ta8}.}
\label{fig2}
\end{figure*}

It can be shown that BSAX and CGRO spectral data imply a
  {\it maximally efficient } acceleration mechanism
 \ci{ta1,ta2}
not dissimilar from what calculated \ci{hoshino}
and observed (e.g., \ci{jager}) in synchrotron nebulae.
The  SSM spectral shape
turns out to be a universal  {\it 2-parameter} function
in the energy range $\sim 1$~keV--1~GeV
[the critical energy $\varepsilon_c $
(proportional to the local magnetic field strength $B_s$
and to $\gamma_*^2$), and
the index $\delta$]. In
the absence of a supra-thermal  component,
 the low-energy ($E \ll \varepsilon_c$)
emission is  dictated by
\syn emissivity of energy index $1/3$.
For intermediate photon energies $E\loe \varepsilon_c$ the spectrum
steepens, and SSM predicts a very distinctive `curvature' of the  
continuum  for specific  combinations  of the relativistic Maxwellian and
power-law components \ci{ta1,ta2}.
The observed  `peak' photon energy $E_p$  is
proportional to the relativistic
 \syn  energy $\varepsilon_c$
(modulo  possible factors due to   Doppler blueshift $\sim \Gamma$,
cosmological
redshift $\sim (1 + z)^{-1}$, and overall IC upscattering, if applicable).
No substantial modification of GRB spectral shapes by inverse Compton (IC)
processes is evident in BSAX and
CGRO data.
 This indicates that IC cooling,
if it occurs at all, `gently' shifts the overall spectrum
dictated by \syn emission.
This is a crucial constraint for many cosmological models characterized
by an overproduction of IC emission compared to pure \syn
(e.g., \ci{mesz3}).
SSM physical parameters can be derived by
 by the combined set of CGRO and  BSAX spectral data \ci{ta8}
\be  10^{3} \loe \gamma_* \loe  10^6 \;\;\;\;\;\;\;\;
1~{\rm G}\loe  B_{s} \loe  10^{3} {\rm G} \en
modulo redshift and IC upscattering factors.
These values are quite natural for optically thin
 synchrotron nebulae powered by MHD outflows.

Interesting deviations from the average low-energy
spectrum are observed in a few cases.
An apparent suppression
of low-energy photons during the {\it initial part of some GRB pulses}
is  detected in a small fraction ($\loe 15 \%$)  of
events  by BATSE \ci{crider}, and GINGA \ci{strohmayer}.
However, difficulties of  the BATSE spectral analysis  at
low energies and intensities,
and uncertainties on the incidence angle of GRBs detected by
GINGA make these measurements somewhat uncertain.
BSAX data can resolve the issue.
Among the  GRBs detected by the WFCs, only 
GRB~960720 \ci{piro5}
(during the first second of the main pulse lasting $\sim 10$~sec
in the GRBM energy band) shows a clear sign of low-energy
suppression.
This time-dependent suppression suggests a modification
from the idealized optically thin conditions assumed in the
simplest version of the SSM.
Low-energy SSM emission below $\sim 50$~keV
 can be temporarily suppressed 
by relativistic plasma effects 
\ci{ta8}.
The fact that  low-energy suppression occurs at the beginning
of GRB pulses is a clear indication that the radiative
environment relaxes from a complex  to a thinner medium
as expected in plasma acceleration models \ci{ta8}.

More surprises may enrich our study of GRBs in the near
future. We need more optical/radio identifications of
reliable GRB counterparts to settle the issue of the 
GRB origin. A systematic  time-resolved spectral analysis is
necessary  for both BSAX and BATSE data
 to test emission models.
 Joint BSAX  and BATSE spectral analysis will be of great
importance for GRBs detected by both instruments.

\vskip .1in

The author thanks all members of the BSAX team for many
lively discussions and exchange of information that 
stimulated  this work.

\end{document}